\documentclass[11pt, prd,preprintnumbers,amsmath,amssymb, superscriptaddress]{revtex4}
 \usepackage[latin1]{inputenc}
\usepackage{amssymb}
\usepackage{amsmath}
\usepackage{enumerate}
\usepackage{graphicx}
\usepackage{subfigure}
\usepackage{multirow}
\usepackage{hyperref}
\usepackage{url}

\def\be{\begin{equation}}
\def\ee{\end{equation}}
\def\ba{\begin{array}}
\def\ea{\end{array}}
\makeatletter

\usepackage{epsf}

\def\be{\begin{equation}}
\def\ee{\end{equation}}
\def\ba{\begin{eqnarray}}
\def\ea{\end{eqnarray}}
\def\beas{\begin{eqnarray*}}
\def\eeas{\end{eqnarray*}}
\def\sla{\raise.15ex\hbox{$/$}\kern-.57em}

\newcommand{\hc}{\text{ h.c.}}
\newcommand{\eV}{\text{ eV}}

\newcommand{\GeV}{\text{ GeV}}
\newcommand{\TeV}{\text{ TeV}}
\newcommand{\pb}{\text{ pb}}
\newcommand{\fb}{\text{ fb}}
\newcommand{\cm}{\text{ cm}}
\newcommand{\mm}{\text{ mm}}

\newcommand{\m}{\text{ m}}

\begin{document}

\title{\Large\bf Displaced Supersymmetry}

\author{Peter W. Graham}
\affiliation{Stanford Institute for Theoretical Physics, Department of Physics, Stanford University, Stanford, CA 94305}

\author{David E. Kaplan}
\affiliation{Department of Physics and Astronomy, The Johns Hopkins University,  Baltimore, MD 21218}

\author{Surjeet Rajendran}
\affiliation{Stanford Institute for Theoretical Physics, Department of Physics, Stanford University, Stanford, CA 94305}
\affiliation{Department of Physics and Astronomy, The Johns Hopkins University,  Baltimore, MD 21218}

\author{Prashant Saraswat}
\affiliation{Stanford Institute for Theoretical Physics, Department of Physics, Stanford University, Stanford, CA 94305}

\preprint{SU-ITP-12/03}


\begin{abstract}
The apparent absence of light superpartners at the LHC strongly constrains the viability of the MSSM as a solution to the hierarchy problem. These constraints can be significantly alleviated by $R$-parity violation (RPV). Bilinear $R$-parity violation, with the single operator $L H_u$, does not require any special flavor structure and can be naturally embedded in a GUT while avoiding constraints from proton decay (unlike baryon-number-violating RPV). The LSP in this scenario can be naturally long-lived, giving rise to displaced  vertices. Many collider searches, particularly those selecting $b$-jets or leptons, are insensitive to events with such detector-scale displaced decays owing to cuts on track quality and impact parameter. We demonstrate that for decay lengths in the window $\sim 1 - 10^3$ mm, constraints on superpartner masses can be as low as $\sim 450 \GeV$ for squarks and $\sim 40 \GeV$ for LSPs.  In some parts of parameter space light LSPs can dominate the Higgs decay width, hiding the Higgs from existing searches. This framework motivates collider searches for detector-scale displaced vertices.  LHCb may be ideally suited to trigger on such events, while ATLAS and CMS may need to trigger on missing energy or multijet signatures.
\end{abstract}

\maketitle

\tableofcontents

\section{Introduction} \label{sec:Intro}
Weak scale supersymmetry is a plausible solution to the hierarchy problem. The simplest implementation of this framework, namely, the Minimal Supersymmetric Standard Model (MSSM), accommodates attractive theoretical features such as gauge coupling unification and candidates for dark matter~\cite{Dimopoulos:1981yj}. This minimal framework has been searched for at various collider experiments including the Large Electron-Positron Collider (LEP), the Tevatron and the Large Hadron Collider (LHC) in the past two decades.  The absence of salient signatures of supersymmetry such as  light superpartners in these experiments places the MSSM in an uncomfortable situation as a solution to the hierarchy problem~\cite{Barbieri:2000gf, Giudice:2006sn, Raby:2007yv,  Dermisek:2009si}. Unlike indirect constraints on the MSSM (such as probes of flavor) that can be accommodated with suitable ultraviolet structures, these collider limits directly constrain the low energy spectrum of the theory. Recent results from the LHC have further motivated exploration of scenarios in which experimental constraints can be satisfied without introducing fine-tuning~\cite{Fan:2011yu,Kribs:2012gx,Brust:2011tb,Papucci:2011wy, LeCompte:2011cn,LeCompte:2011fh,Feng:2011aa,Allanach:2012vj,Nikolidakis:2007fc, Csaki:2011ge}.

It is useful to ask if simple extensions of the MSSM can still function as a viable solution to the hierarchy problem. This question is of great experimental relevance, as the extraction of new physics from significant standard model backgrounds in the collider environment depends on more than just the production of the new particles. The subsequent interactions of the produced particle, including its potential decays into the standard model, have to be specified and searches are tailored for these specific scenarios. In the absence of dedicated searches probing these decay topologies, we may be unable to discover the new physics despite producing it copiously. The identification of new possible decay topologies therefore provides an important direction for experimental searches.

$R$-parity violation (RPV) is one such simple extension of the MSSM. Introduced initially as a simple parity to forbid proton decay, $R$-parity ensures the stability of the lightest supersymmetric particle (LSP).  While a weak scale neutral LSP can naturally explain the observed dark matter abundance, it results in large missing energy in supersymmetric collider events. The most stringent constraints on the MSSM arise from searches for such events. In the presence of RPV, causing the LSP to decay to visible standard model particles, the sensitivity of these missing energy searches to supersymmetry events can be eliminated or significantly decreased, allowing for the existence of light superpartners~\cite{KaplanUDD,Carpenter:2008sy}. $R$-parity violation must be introduced with care as the stability of the proton requires that either baryon or lepton number violating operators be small. Na\"{i}vely, the relative suppression necessary between low energy baryon and lepton number violation is in conflict with the demands of a unified theory at the GUT scale where both these operators would naturally be expected to arise at the same order. Since the apparent unification of gauge couplings due to weak scale superpartners is a strong motivation for the MSSM, it is desirable to consider violations of $R$-parity that can naturally be embedded into a GUT framework.

Bilinear $R$-parity violation, in which $R$-parity is broken through terms of the form $W \supset \mu_L L H_u$, is a natural framework to accommodate these ideas~\cite{Hall:1983id}. In this work we show that if the bilinear RPV couplings are small enough that the decays of the LSP are displaced, then supersymmetry with a relatively light spectrum of superpartners could be hidden from current LHC searches. We review this framework in sections~\ref{sec:synopsis} and~\ref{sec:BRPV} where we highlight the key theoretical and experimental points that allow displaced decays from bilinear RPV to address the constraints on the MSSM.  We discuss collider searches for displaced decays in section~\ref{sec:DisplacedCollider}. We give constraints on this framework in sections~\ref{sec:ChiLSP} and~\ref{sec:Constraints} for various choices of LSP, and show that light squark masses are allowed when the $R$-parity violation is weak enough to give macroscopically displaced decays of the LSP. In section~\ref{sec:Higgs}, we discuss new decays of the Higgs boson in this model and the possibility of hiding a light Higgs. In section~\ref{sec:BRPVLowEnergy}, we consider indirect limits from neutrino masses and proton decay on these RPV operators. Finally, in section~\ref{sec:discovery}, we discuss prospects for discovering displaced decays from bilinear RPV.  

\section{Synopsis}
\label{sec:synopsis}

It is well known that RPV can hide SUSY by removing the normal missing energy signatures. However this usually requires a careful choice of RPV operators to avoid generating proton decay and other problematic operators and can therefore be difficult to embed in a grand unified theory (GUT) (see e.g.~\cite{Barbier:2004ez}). For example, the RPV operator $UDD$ can reduce many of the constraints on SUSY, but must be generated without the L-violating operators that naturally come with it in a unified theory and must also have a complicated flavor structure to avoid all constraints (see e.g.~\cite{Nikolidakis:2007fc, Csaki:2011ge} for recent work using the MFV hypothesis).  Instead, we consider a model that is simple, easily embeddable in a GUT and allows a natural, light superpartner spectrum. We show how this can give long decay lengths for the LSP that hide supersymmetry from existing searches. 

In addition to the usual MSSM, we add a hidden sector that breaks $R$-parity. As we discuss below, this essentially adds only the lowest dimension RPV operator to the superpotential, the bilinear $W \supset \mu_L L H_u$. The other RPV operators are significantly suppressed and mostly irrelevant for the phenomenology. This removes the proton decay problem that comes with adding all RPV operators since this model violates lepton number but not baryon number.  This model can arise naturally from a GUT without reintroducing proton decay because of doublet-triplet splitting, as discussed in Section~\ref{sec:BRPV}.

This single addition to the MSSM can completely alter the phenomenology, greatly weakening bounds on superpartner masses. In certain parts of parameter space, in particular when the LSP is a neutralino, all of the squarks can be as light as $\sim 450 \GeV$. This occurs because of the dilution (though not elimination) of missing transverse momentum (MET) when the LSP decays, combined with the fact that the leptons and/or heavy-flavor jets from the decay will not be identified in most analyses if their tracks do not point back to the interaction vertex. The long lifetime of the LSP is crucial in ensuring the latter~\footnote{Previous work~\cite{AristizabalSierra:2008ye, deCampos:2008re} has considered the case of neutralino LSPs with displaced decays, focusing on a particular parameter space where RPV operators generate neutrino masses; here we consider more general LSPs and discuss the parameter space that allows for light colored superpartners with current constraints.}. This model provides a well-motivated example of displaced decay phenomenology (also a feature of hidden-valley scenarios~\cite{Strassler:2006ri, Strassler:2006im}). There are many possible choices for the LSP that can achieve a light spectrum; we discuss the phenomenology of these different parts of parameter space in section~\ref{sec:ChiLSP} and~\ref{sec:Constraints}. Although this model can hide supersymmetry from existing searches, appropriately designed searches selecting displaced vertices or kinked tracks can have significant discovery potential due to the low Standard Model backgrounds.

If the LSP is light enough, then displaced vertices can also arise in events with a Higgs boson decaying to two LSPs. This can be the dominant Higgs decay mode, rendering most collider searches insensitive and allowing a Higgs as light as 82 GeV (similar to~\cite{KaplanUDD} though with a different RPV scenario, see also~\cite{Bellazzini:2010uk, Luty:2010vd, Falkowski:2010cm, Banks:2008xt, AristizabalSierra:2008ye, Chang:2008cw, Chang:2007de, Dermisek:2006wr}). In this scenario the little hierarchy problem is eliminated and SUSY can be completely natural. The Higgs can also have a subdominant branching ratio to neutralino LSPs, in which case it may be observed in both Standard Model channels and in displaced vertex events. 

\section{The Model: Bilinear \emph{R}-parity Violation}\label{sec:BRPV}

The most general superpotential for the MSSM fields contains possible baryon number and lepton number violating operators. The baryon and lepton conserving part of the superpotential includes the Higgs $\mu$ term and Yukawa interactions:
\be
W_{0}=\mu H_u H_d + y^u_{ij} H_u Q_i U_j + y^d_{ij} H_d Q_i D_j + y^e_{ij} H_d L_i E_j
\ee
The only allowed renormalizable operator violating baryon number is 
\be
W_{\Delta B \neq 0}=\frac{1}{2} \lambda^{''}_{ijk} U_i D_j D_k
\ee
while the lepton-number violating operators include
\be
W_{\Delta L \neq 0}=\frac{1}{2} \lambda_{ijk }L_i L_j E_k + \lambda^{'}_{ijk} L_i Q_j D_k +\epsilon_{i} L_i H_u
\ee
where the indices $i,j,k$ run over the three generations of fermions.

Many of these parameters are strongly constrained by low-energy observables. In particular, the presence of both baryon and lepton number violation leads to proton decay $p \rightarrow e^+\pi^0$ at a rate proportional to $\sum_{i=2,3} |\lambda^{'}_{11i} \lambda^{''}_{11i}|^2$, suppressed only by the squark masses rather than the unification scale. To avoid this it is standard to impose the symmetry of matter parity (equivalently $R$-parity), $(-1)^{3(B-L)}$, forbidding all of the baryon and lepton number violating terms. This precise symmetry however is not forced upon us by the data. Imposing either baryon or lepton number alone is sufficient to prevent proton decay; more generally it is technically natural for the coefficients of some or all of the superpotential terms to be arbitrarily small. In this paper we consider $R$-parity violating operators consistent with all constraints that nevertheless can have drastic implications for searches for supersymmetry.

In particular we focus on the simple and predictive scenario of bilinear $R$-parity violation (BRPV), where we assume that in some basis the dominant $R$-parity violating operators are the superpotential bilinears
\be
W \supset \mu_{L,i} L_i H_u
\ee 
which are rotated to lepton-number-violating trilinears in the mass basis:
\be
W \supset \epsilon_i y^d_{jk} L_i Q_j D_k + \epsilon_i y^e_{jk} L_i L_j E_k 
\label{eq:ModelTrilinear}
\ee 
where $\epsilon_i = \frac{\mu_{L,i}}{\mu}$ are the lepton-Higgs mixing angles (we assume $\mu_{L_i} \ll \mu$). In this paper we work with the superpotential of equation~\ref{eq:ModelTrilinear}, with all other trilinear terms assumed to have much smaller coefficients than the above. Note that this gives a predictive structure, implying that $R$-parity violating processes will dominantly involve fermions of the third generation, in particular the bottom quark. 

This model can be achieved if the $R$-parity violation is communicated to the MSSM through a hidden sector, for example if $R$-symmetry is broken by the vev of some gauge-invariant operator $\mathcal{O}$ that couples to the standard model at a scale $\Lambda$ (see e.g.~\cite{Numass}). If $\mathcal{O}$ has dimensions $n$ and vev $M^n$, we generate for the bilinear term (ignoring generation indices) 
\be
\frac{\mathcal{O}}{\Lambda^{n-1}}L H_u \to \frac{M^n}{\Lambda^{n-1}}L H_u.
\ee
The resulting $L$-$H$ mixing angle is $\epsilon = \frac{M^n}{\Lambda^{n-1} \mu}$. As $\Lambda$ is increased this can become arbitrarily small, while remaining larger than the coefficients of the trilinear terms induced by direct couplings to $\mathcal{O}$ by a factor $\frac{\Lambda}{\mu}$.

Bilinear $R$-parity violation can naturally be embedded in a UV theory with $SU(5)$ gauge and matter unification. To generate trilinear RPV terms in an $SU(5)$ invariant way one would write the operator $\bar{\mathbf{5}}_L\bar{\mathbf{5}}_L\mathbf{10}$, which generates all three types of trilinear terms at an equal level. To avoid this one must introduce$SU(5)$ breaking. This can be done in a minimal way by considering a bilinear RPV term. The $SU(5)$ invariant operator $\mu_{L,i} \bar{\mathbf{5}}_L\mathbf{5}_{H_u}$ includes both $L_i H_u$ and $D_i H^C_u$ terms, where $H^C$ denotes colored Higgs triplets, and thus also violates both baryon and lepton number. However, any phenomenologically viable $SU(5)$ theory must decouple the $H^C$ at low energies in order to satisfy proton decay constraints (its mass $M_{H^C}$ must be on the order of the unification scale or greater if dimension-five operators are generated by the color triplet higgsinos). Achieving this doublet-triplet splitting requires nontrivial model-building, but once accomplished the baryon number violation from $\mu_{L,i} \bar{\mathbf{5}}_L\mathbf{5}_{H_u}$ is suppressed by the mass of the colored Higgs if it is present in the spectrum (and can be even further suppressed in orbifold models~\cite{Kawamura:2000ev,Hall:2001pg}). In the mass basis, the resulting $UDD$ trilinear term is 
\be
W_{\Delta B \neq 0}= \frac{\mu_{L,i}}{M_{H^C}} y^d_{jk} D_i D_j U_k.
\ee
This baryon number violation can easily be small enough to satisfy proton decay constraints (section~\ref{sec:pdecay}). 

Below the scale that supersymmetry breaking is communicated to the visible sector, SUSY-breaking scalar mixing terms between the $L_i$ and Higgses will be generated. These mixings introduce off-diagonal gaugino couplings, leading to tree-level neutrino masses and mixing of the gauginos and leptons. This can affect the decay of neutralino LSPs, as discussed in section~\ref{sec:ChiLSP}, but has negligible effect in the case of a sfermion LSP. 

\begin{table}[b]
\caption{Dominant decay modes of some possible LSPs with bilinear $R$-parity violation.}
\begin{center}
\begin{tabular}{| l | c | c | c | c | c |}
\hline
\textbf{LSP} & $\tilde{\chi}$ & $\tilde{\nu}$ & $\tilde{\tau}$ & $\tilde{u}_L$ & $\tilde{b}$\\ \hline
\textbf{Dominant Decays} & $\nu b\bar{b}$, $\nu \tau l$ & $b\bar{b}$ & $l^\pm\nu$ & $l^{\pm}q$ & $b\nu$ \\ \hline
\end{tabular}
\end{center}
\label{tab:decaymodes}
\end{table}

\section{Collider Searches for Displaced Vertices}
\label{sec:DisplacedCollider}

Existing constraints on $R$-parity violation with the lepton-number violating operators $LQD$ and $LLE$ require sfermion LSPs to be heavier than $\sim$ 80 GeV assuming prompt decays (see e.g. \cite{ALEPH4jet}). At hadron colliders, searches in final states with missing energy, heavy quarks and/or charged leptons can place upper bounds on the squark and gluino masses, thereby testing SUSY naturalness. 

However, the decay length of the LSP depends on the unknown dimensionless parameters $\epsilon_i$, which can naturally be  small. Neutrino mass constraints in general require $\epsilon_i \lesssim 10^{-3}$ (section~\ref{sec:numass}); LSP decays are additionally suppressed by Yukawa couplings and, in the neutralino case, by three-body phase space as well. For sufficiently small $\epsilon_i$ ($\sim 10^{-3}$ for neutralinos, $\sim 10^{-5}$ for sfermions) the LSPs may decay at macroscopic distances ($\gtrsim \mm$) from the beamline in collider experiments.

Most collider searches that use information about charged tracks place track quality cuts that include requirements on the track impact parameter (the closest approach of the extrapolated track to the beamline or primary vertex). These cuts typically require impact parameters less than $O(\text{cm})$ at the LEP detectors ALEPH, DELPHI and OPAL~\cite{ALEPHdetector,DELPHIdetector,OPALtracks}, and $O(\text{mm})$ at CDF, D0, ATLAS and CMS (see e.g.~\cite{ATLASdetector,CMStrack,CDFHV,CDFleptontrack2006,D0leptontrack2007}). For decay lifetimes significantly longer than these minimum impact parameters, tracks from displaced decays will usually fail these cuts, if they are reconstructed at all. Electrons and muons in particular are required to have tracks with small impact parameter at CMS~\cite{CMSElectronReconstr, CMSMuonReconstr} and in most searches at ATLAS~\cite{ATLASelectron0,ATLASelectron, ATLAS1lep, ATLAS2lep, ATLASDilepRes, ATLASIdentFlavLep, ATLASsamesigndilep}, so the presence of charged leptons in LSP decay does not guarantee sensitivity. Similarly, impact-parameter-based tagging of heavy quarks or hadronically decaying taus requires tracks with impact parameter less than $\sim \text{mm}$ at ATLAS and CMS \cite{ATLASbtag,CMSbtag,ATLAStautag}, so these objects will not be tagged if they originate from a displaced vertex. At CMS, even ordinary jets are reconstructed using particle-flow techniques, and CMS analyses generally require all jets to have at least one well-measured charged track~\cite{CMSjetquality2009}. Therefore we assume that CMS supersymmetry searches do not place constraints in the case of displaced LSP decay.  

Decays of slepton and neutralino LSPs through bilinear RPV can produce final states with neutrinos. However, the missing transverse energy in such events can be suppressed compared to the usual assumption of a stable neutralino LSP. We will find that for some regions of parameter space, squarks with mass as low as $\sim 450 \GeV$ for all three generations (corresponding to production cross-sections of $\approx 2 \pb$) can avoid constraints from existing searches in jets and MET, regardless of whether the LSP decay is displaced or not. However, due to the presence of bottom quarks or leptons in these LSP decays, searches in missing energy and leptons or $b$-tagged jets can place stronger constraints when the LSP decay is prompt. A light spectrum therefore requires LSP decay lengths $\gtrsim \mm$.

\begin{table}[b]
\caption{For various collider searches applicable to $R$-parity violating supersymmetry, we summarize the factors that limit the sensitivity in the case when the LSP decay is displaced. These are discussed in more detail in section~\ref{sec:DisplacedCollider}.}
\begin{center}
\begin{tabular}{| p{5 cm} | p{12cm} |}
\hline
\textbf{Search} & \textbf{Factors limiting sensitivity to displaced decays from RPV}  \\ \hline
ATLAS jets + MET & $R$-parity violation gives suppressed missing energy \\ \hline
CMS jets + MET & Events containing jets without well-measured tracks are rejected \\ \hline
ATLAS/CMS b-jets + MET & $b$-tagging algorithm requires track impact parameter $< 1-2 \mm$\\ \hline
ATLAS/CMS leptons & Lepton reconstruction usually requires track impact parameter $< 1-2 \mm$ \\ \hline
LEP searches & Track impact parameters must be $\lesssim$ few cm \\ \hline
ATLAS displaced vertex search & Requires muon with $p_T > 45 \GeV$ to trigger \\ \hline
 \end{tabular}
\end{center}
\label{tab:avoid}
\end{table}

Even if the final state contains no invisible particles, an event with a highly displaced decay may appear to have missing transverse momentum if objects from displaced decays are incorrectly assumed to originate from the primary vertex. The fractional mismeasurement of the momentum of an object produced at a displaced vertex will scale roughly as the decay lifetime over the detector size for small displacements. Therefore we expect this effect to be relevant mostly for lifetimes $\gtrsim 10 \cm$. A more precise statement would require detailed simulation of detector performance and object reconstruction in response to highly displaced decays, which we do not attempt here (see however~\cite{Fan:2012jf}, in which this effect is studied at the geometric level and found to be negligible for lifetimes $\lesssim 10\cm$). Throughout the rest of this paper we assume that the apparent missing momentum is negligible in any event that is accepted as signal in collider analyses using MET information.  

A limited number of analyses explicitly relax track quality cuts and search for displaced vertices and large impact parameter tracks. In the following section we describe the various searches that have sensitivity to displaced decays.

In estimating the constraints on the superpartner spectrum from searches at hadron colliders we simulate events using PYTHIA 6 \cite{PYTHIA} and reconstruct jets using FastJet~\cite{Fastjet}. For processes with three-body neutralino decays, the events are weighted to account for the nontrivial matrix element of the neutralino decay. We use Prospino2~\cite{Prospino1, ProspinoSquark, ProspinoStop} to determine NLO cross-sections for the production of squarks and gluinos. We emphasize that our quantitative results are only estimates, as we neglect several effects including detector simulation, parton shower-matrix element matching for radiation, etc. However, due to the rapid variation of the colored superpartner production cross-sections as a function of mass, errors in our determination of cross section times acceptance times efficiency will generally give small corrections to the mass bounds on squarks and gluinos.

\subsection{Displaced jets}
\label{sec:DisplacedJets}

The CDF, D0 and ATLAS collaborations have published searches~\cite{CDFHV,D0displacedjet, ATLASdisplacedjet, ATLASMS} for displaced vertices with many associated tracks, such as would be produced by long-lived particles decaying to one or more jets.

The ATLAS collaboration has presented a search for displaced vertices in events passing a muon trigger, using $33 \pb^{-1}$ of data~\cite{ATLASdisplacedjet}. Events are required to have a muon with $p_T > 45 \GeV$ visible in both the inner detector and the muon spectrometer, though no requirement is made on the impact parameter of this track. At least one displaced vertex is required with transverse impact parameter $4 \mm < |d| < 180 \mm$. At least four outgoing tracks are required to be associated with the displaced vertex, greatly reducing sensitivity to displaced decays involving only leptons. The results of ~\cite{ATLASdisplacedjet} are interpreted as bounds on simplified models with degenerate squarks and a neutralino LSP decaying to a muon and two jets. To estimate bounds for other decay modes we take the excluded cross-section in these models and correct for the different muon cut acceptance (see section~\ref{sec:ChiLSP}). 

The ATLAS collaboration has also performed a search for a Higgs boson decaying to long-lived, weakly-interacting particles that can decay hadronically in the muon spectrometer or the outer regions of the hadronic calorimeter~\cite{ATLASMS}. A dedicated trigger~\cite{ATLASLLTrigger} based on identifying multiple muon ``regions of interest" isolated from any calorimeter jets or inner detector tracks is used to trigger on such events. This analysis excludes lifetimes $c\tau$ for the long-lived particles of $\sim .5-1.5 \m$ for Higgs masses in the range $120-140$ GeV and long-lived particle masses in the range $20-40$ GeV. This search rapidly loses sensitivity for boosted decay lengths $\lesssim \m$ due to the exponentially small probability for the long-lived particle to reach the muon spectrometer. 

The CDF and D0 collaborations have performed searches~\cite{CDFHV,D0displacedjet} for a Higgs decaying into a pair of neutral long-lived particles that decay in the tracker to produce displaced vertices with jets (including some heavy-flavor jets). These searches can apply directly to our scenario of a Higgs decaying to superpartners, but the limits they place on the Higgs cross-section times branching ratio are greater than the Standard Model total production cross-section and thus do not constrain models without enhanced Higgs production. Likewise, these searches will likely not constrain supersymmetry events with displaced decays if the total SUSY production cross-section is $\lesssim 1 \pb$, corresponding to squark and gluino masses $\gtrsim 300 \GeV$.         

The ALEPH experiment has performed a search~\cite{ALEPHstops} for stops nearly degenerate with a neutralino LSP, allowing for the stop decay length to be macroscopically long. This analysis included a search for stops decaying in the tracking volume, producing displaced vertices with multiple associated tracks. However, the selection for this search requires the event to have visible energy less than $.2\sqrt{s}$, which will not occur in pair-production events for any of the LSPs we consider.

\subsection{Displaced leptons}
\label{sec:DisplacedLeptons}

The ALEPH, D0 and CMS collaborations have published searches~\cite{ALEPHsleptons,ALEPHgmold,ALEPHgmnew,D0dielectron,D0dimuon,CMSdilepton} for displaced vertices producing charged leptons.

The ALEPH collaboration has performed searches in LEP2 data for tracks with kinks or large impact parameters~\cite{ALEPHsleptons,ALEPHgmold,ALEPHgmnew}. These searches require that there be no more than a few (2--6) charged tracks in an event, so they will only apply when neither of the LSPs decay to jets within the tracking volume, i.e. when both decay through the $LLE$ operator. The combination of the searches for kinks and for large impact parameters is $O(1)$ efficient for charged particles with decay lengths $\sim 1 - 3000 \cm$~\cite{ALEPHsleptons}. A total of $628 \pb^{-1}$ at center-of-mass energies $\sim 200 \GeV$ was analyzed~\cite{ALEPHgmnew} in these searches. At 95\% CL the bound on average cross-section times branching ratio times efficiency is $6 \fb$. 

The CMS collaboration has performed a search~\cite{CMSdilepton} in $1.1 \fb^{-1}$ for opposite-sign dileptons (electrons and muons) arising from a displaced vertex. The leptons are required to have $p_T > 38 \GeV$ (25 GeV) for electrons (muons). Because the two leptons are assumed to be the only decay products of the long-lived particle, the momentum vector of the dilepton and the vector from the primary vertex to the displaced vertex are required to be separated in azimuthal angle by less than .8 (.2) radians for electrons (muons). For a dimuon vertex the muons are further required to be separated by $\Delta R > .2$ to ensure good trigger efficiency. The search is interpreted in terms of a heavy ($> 200 \GeV$) Higgs decaying to two long-lived neutral particles, at least one of which decays to two leptons. The leptonic decay of a neutralino to $\nu \tau l$ can give a similar topology when the tau decays leptonically. Since neutralinos from squark decay are highly boosted, the visible momentum of their decay products will be aligned with the original neutralino momentum despite the presence of neutrinos in the decay. 

The D0 collaboration has published searches for opposite-sign muons~\cite{D0dimuon} and electrons~\cite{D0dielectron} arising from a displaced vertex; these place weaker constraints than the CMS search~\cite{CMSdilepton}.

\section{Neutralino LSPs}
\label{sec:ChiLSP}

In this section, we show that a neutralino LSP with lifetime $c\tau$ between $\sim \mm$ and $\sim \m$ significantly changes the collider signals of supersymmetry, removing many current constraints. With the $R$-parity violating superpotential term $L_i H_u$, a neutralino LSP can undergo three-body decays to $\nu b \bar{b}$ (through $LQD$) or to $\nu \tau l$ where $l$ is a charged lepton (through $LLE$). The branching ratios depend on the neutralino couplings and the squark and slepton masses; if all sleptons are degenerate then the $\nu b \bar{b}$ mode dominates, though if the right-handed stau is sufficiently light then $\nu \tau l$ can be significant or dominant. 

SUSY-breaking mixings of the sleptons and Higgs scalars can introduce decay modes for the neutralino beyond what the supersymmetric bilinear allows, including two-body decays to a neutrino and a $Z$ or a Higgs. We will assume that such decays are suppressed, either because they are not kinematically allowed or because the coefficients of the $\tilde{l} h$ mixing terms are sufficiently small. In section~\ref{sec:numass} we discuss the suppression of these terms in models with low-scale SUSY breaking.

The production cross-section of neutralinos at colliders depends on the neutralino mixing matrix and the masses of other superpartners, so there is no model-independent direct bound on the mass of a neutralino LSP. We therefore focus on the bounds on squark and gluino masses in the light neutralino scenario. We will consider a simplified spectrum with three generations of degenerate squarks, a gluino, and a 50 GeV neutralino.

\subsection{Jets + MET searches and short lifetimes}
\label{sec:JetsMETshort}
\begin{figure}
\includegraphics{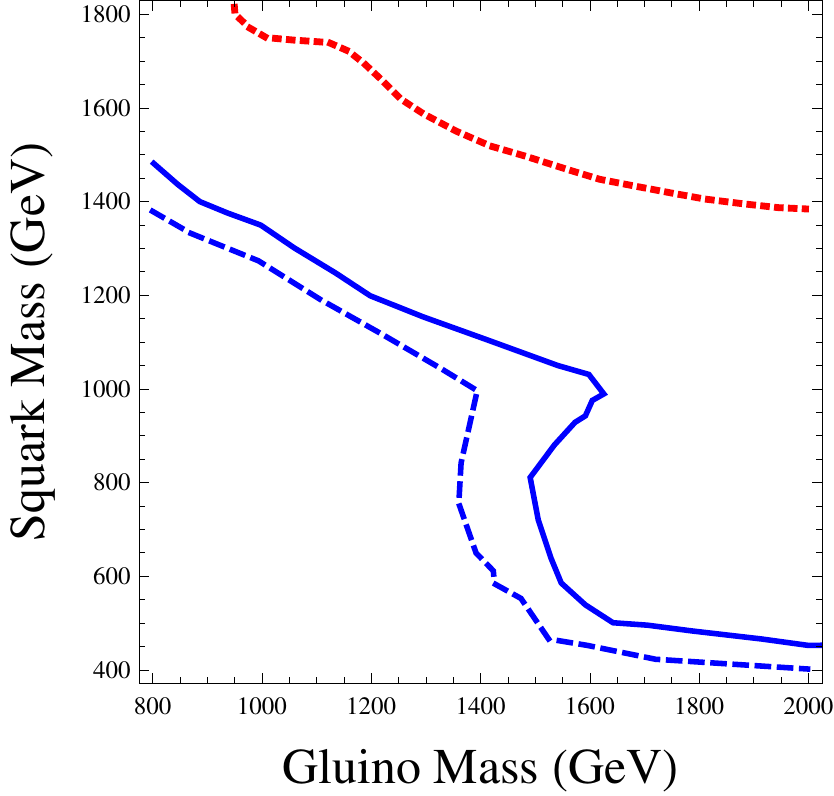}
\caption{Constraints on squark and gluino masses assuming a light neutralino LSP with all other sparticles decoupled, for neutralino lifetimes $1 \cm \lesssim c\tau \lesssim 10 \cm$. The lower, blue contours indicate the bounds for displaced decays of a 50 GeV neutralino to $\nu b \bar{b}$, assuming all squarks are degenerate. The solid blue line indicates the contour where the cross-section times acceptance for some search, calculated using the central values of renormalization and factorization scales in Prospino and ignoring detector effects, equals the excluded cross-section times acceptance times efficiency reported by ATLAS. The dashed blue line lines give the contour where the calculated cross-section times acceptance is 1.3 times the exclusion; this reflects the possible effects of detector efficiency and uncertainties in the theoretical cross-sections and the detector response. The dotted red line is the bound in the case of a stable massless neutralino, with the third generation of squarks taken to be heavy and decoupled (from the ATLAS jets + MET search in $4.7\fb^{-1}$~\cite{ATLASjetsMET5}).
}         
\label{fig:ChiSquarkGluino}
\end{figure}

The three-body neutralino decay can suppress the missing energy enough to evade searches relying on jets and MET. For decay lengths $\lesssim 10 \cm$, the displaced origin of the jets from neutralino decay will not significantly affect the measurement of their momenta, so we may determine the signal efficiencies of LHC searches assuming that all jet momenta are correctly determined. The blue (solid and dashed lines) contours of figure~\ref{fig:ChiSquarkGluino} show the bounds on squark and gluino masses from ATLAS searches in jets and missing energy~\cite{ATLASjetsMET5,ATLASjetsMET1.04,ATLASjetsMET35,ATLASmultijet5,ATLASmultijet1.3} for a 50 GeV neutralino with a displaced decay through a sneutrino to $\nu b \bar{b}$. Shown for comparison are the bounds from~\cite{ATLASjetsMET5} on the MSSM without RPV (red dotted line), for a spectrum with only the squarks of the first two generations, the gluino, and a stable massless neutralino. We find that displaced decays of the neutralino can allow the squarks of all three generations to be as light as $\sim 450 \GeV$. This requires a heavy gluino ($\gtrsim 1500 \GeV$) however, as squark-gluino and gluino-gluino events have much higher acceptance for searches in final states with many (6+) jets. 

If the neutralino decays to $\nu \tau l$, then the resulting leptons will not satisfy the track impact parameter cuts imposed on signal leptons in most LHC searches. Events in zero-lepton analyses will be vetoed however if they contain an isolated muon \emph{candidate}, which need not meet these quality cuts. We assume that electrons that are not identified as such due to their displacement are reconstructed as jets. Decays to $\nu \tau l$ do however have enhanced missing energy compared to decays to $\nu b \bar{b}$, increasing the signal efficiency. We find that the net result is that the lower bound on the squark masses remains similar to the $\nu b \bar{b}$ case, $\sim 450 \GeV$.

Since running from a high scale tends to draw squark masses up rapidly if the gluino is heavy, nontrivial model-building is necessary to achieve a spectrum with a $\gtrsim 1.5 \TeV$ gluino and $\sim 450 \GeV$ squarks. If at the SUSY breaking scale the squarks have zero mass and the gluino has mass 1.6 TeV, then in order for the physical squark masses to be $\sim 500 \GeV$ the scale of SUSY breaking must be $\lesssim 5 \TeV$ (possible in models such as deconstructed gauge mediation~\cite{Deconstructed}). An alternative to such low-scale SUSY breaking is to give the gluino a Dirac mass, in which case its contribution to the squark masses is no longer logarithmically divergent~\cite{Supersoft,Brust:2011tb} and a larger squark-gluino hierarchy can be maintained.     

\begin{figure}
\includegraphics{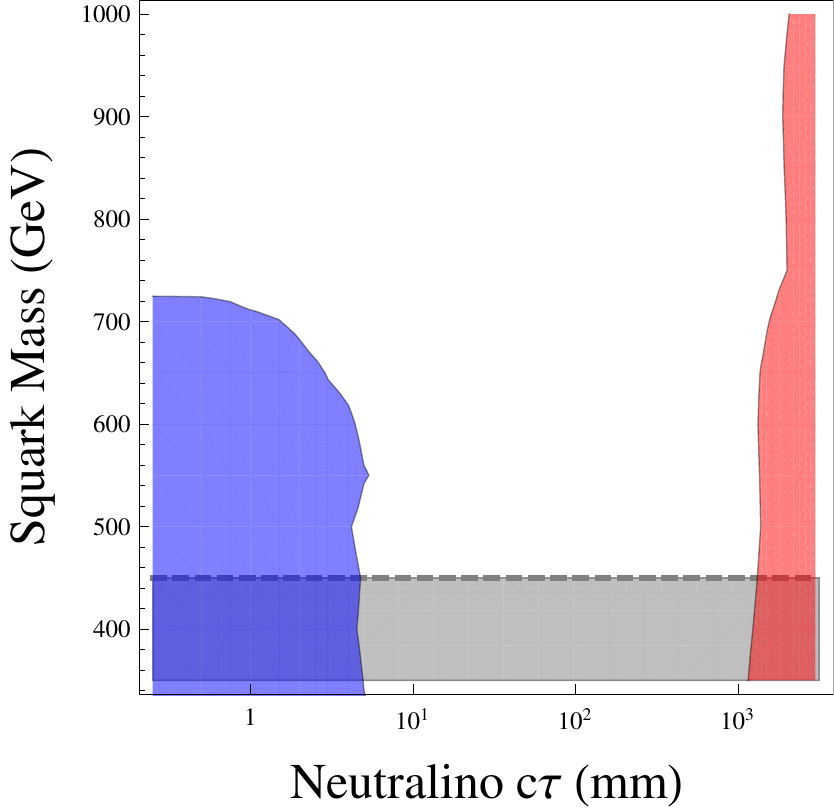}
\caption{Constraints in the plane of squark mass and neutralino decay lifetime for a spectrum with degenerate squarks, a 2 TeV gluino and a 50 GeV neutralino decaying to $\nu b \bar{b}$. Short decay lengths (blue region) are constrained by searches in $b$-tagged jets and missing energy (section~\ref{sec:bJetsMET}). Long decay lengths (pink region) are constrained by jets + MET searches assuming that neutralinos look like missing energy if they decay outside the calorimeter (section~\ref{sec:JetsMETlong}). For decay lengths $\lesssim$ 10 cm, for which we can assume that the error in momentum reconstruction of displaced objects is small, searches in jets + MET require the squarks to be heavier than $\sim 450 \GeV$ (section~\ref{sec:JetsMETshort}). For larger decay lengths the bounds from this search depend on the detector response to and reconstruction of displaced and/or trackless jets, which we do not model. If we make the approximation of extending the prompt decay jets + MET bound of 450 GeV to larger decay lengths then the gray region in the figure is excluded.
}         
\label{fig:ChiDecayLength}
\end{figure}

\subsection{Jets + MET searches and long lifetimes}
\label{sec:JetsMETlong}
For longer decay lengths, the efficiency of these searches depends on how jets from highly displaced vertices are reconstructed, as discussed in section~\ref{sec:DisplacedCollider}. Independent of the detector response to displaced decays, we can determine upper bounds on the decay length by considering events in which both neutralino LSPs escape the detector, so that the neutralino appears stable. We assume that neutralinos will be invisible if they decay outside the hadron barrel calorimeter (even if the decays produce signals in the muon systems). The resulting constraints from the ATLAS searches in jets and missing energy are shown in pink in figure~\ref{fig:ChiDecayLength} in the plane of squark mass and neutralino proper decay length ($c \tau$); in general decay lengths $\gtrsim 1 \m$ are constrained. 

\subsection{Leptons + MET and \emph{b}-jet + MET searches}
\label{sec:bJetsMET}
If the neutralino decays promptly, then the resulting bottom quarks and/or leptons provide sensitivity to existing collider searches. If the neutralino decays through an off-shell sneutrino to $\nu b\bar{b}$, then ATLAS searches in final states with $b$-tagged jets and missing energy (in up to $2.05 \fb^{-1}$~\cite{ATLASbjet2,ATLASbjet830,ATLASbjet36}) are on the brink of excluding the light squark parameter space. If we assume that each $b$ in a jet gives a 60\% probability of the jet being $b$-tagged, we find that the signal efficiency decreases with mass such that even $\sim 350 \GeV$ squarks are just barely unconstrained by the $2 \fb^{-1}$ search if the gluino is $2 \TeV$, though $5\fb^{-1}$ is expected to exclude squark masses up to $\sim 650 \GeV$. 

If the dominant decay is through a right-handed stau to $\nu \tau l$, then an ATLAS search in final states with two leptons~\cite{ATLAS2lep} requires the squarks to be heavier than $\sim 850 \GeV$ for prompt decays.

A neutralino with a decay length greater than a few millimeters, however, avoids such searches, as its decay products will not satisfy the track impact parameter requirements for lepton reconstruction and heavy flavor tagging. In figure~\ref{fig:ChiDecayLength} we plot the region in the squark mass-neutralino lifetime plane that can be constrained by current ATLAS searches in $b$-jets plus MET. We approximate the impact parameter of tracks in a jet from neutralino decay by the ``impact parameters" of the $b$ quark progenitors of the shower, and assume that a jet is rejected by the $b$-tag algorithm if all of these $b$ quarks fail the impact parameter cuts of~\cite{ATLASbtag}. Since a finite neutralino lifetime may increase the probability for the resulting jets to be flavor-tagged, we conservatively assume that a jet from neutralino decay has a 100\% probability of being $b$-tagged if it is not rejected on the basis of impact parameters. Figure~\ref{fig:ChiDecayLength} shows that the resulting bounds from $b$-jets searches cut out for lifetimes above $\sim 5 \mm$.

\subsection{Displaced vertex searches}
\label{sec:ChiDispVert}
The searches for displaced decays at the LHC that are relevant to displaced BRPV decays include the CMS search for displaced dileptons~\cite{CMSdilepton}, an ATLAS search for displaced vertices~\cite{ATLASdisplacedjet}, and an ATLAS search for decays in the muon spectrometer~\cite{ATLASMS}. These were described in section~\ref{sec:DisplacedCollider}.

The CMS search for displaced dilepton vertices~\cite{CMSdilepton} will be sensitive to neutralino decays to $\nu \tau l$ in which the tau decays leptonically. We roughly estimate the bounds on our scenario by assuming that the efficiency to pass the cuts on the lepton momenta (including the $p_T$ cuts, the cut on azimuthal separation of the dilepton momentum and displacement vector, and the cut on $\Delta R$ between muons) can be factored out of the total signal efficiency times acceptance. We then multiply the cross-section exclusions presented in~\cite{CMSdilepton} for a 400 GeV Higgs decaying to two 50 GeV long-lived particles by the relative efficiency of these cuts in the Higgs production and displaced SUSY scenarios. The dimuon channel gives the stronger bounds; for 450 GeV squarks we find that the branching ratio of the neutralino to $\nu \tau \mu$ is constrained to be $\lesssim 3\%$. If the $\epsilon_i$ are equal and the sbottoms and sleptons are degenerate, then this branching ratio varies from $3-7\%$ depending on the neutralino composition. However, the branching ratio to leptons can be suppressed if $\epsilon_\tau$ is larger than the other $\epsilon_i$ and/or if the right-handed sleptons are heavier than the others.

The ATLAS search for displaced vertices~\cite{ATLASdisplacedjet} selects events with both a high-$p_T$ muon and a displaced vertex with four or more outgoing tracks. For neutralino decays to $\nu \tau l$, the latter condition will be satisfied only if one or more taus undergo three-prong decays (a 15\% branching ratio). We roughly estimate the bounds this search places on our scenario by comparing to the bounds reported in ~\cite{ATLASdisplacedjet} for a model with squarks degenerate at 700 GeV, a neutralino LSP at 108 GeV and all other particles decoupled.  We assume that the signal acceptance times efficiency can be expressed as a product of the efficiency of the muon $p_T$ cut and the efficiency to identify and accept a displaced vertex. In events with exactly one neutralino decaying leptonically, we assume that the latter is half as large as for the case studied in~\cite{ATLASdisplacedjet}. With these assumptions we find that this search does not place any bound on the neutralino branching ratios for squark masses greater than $\sim 400 \GeV$.

The displaced vertex results of~\cite{ATLASdisplacedjet} use only $33 \pb^{-1}$ of data. Because the search is essentially zero background (the 90\% CL upper limit on the background is .03 events in $33 \pb^{-1}$), additional data can rapidly improve the limits placed by this analysis. For example, if zero events were seen in $1 \fb^{-1}$ of data, then with the same assumptions as above we find a bound on the neutralino branching ratio to $\nu \tau \mu$ of $\sim 5\%$ for a squark mass of 450 GeV, comparable to the constraints from the CMS search for displaced dileptons~\cite{CMSdilepton}. 

The ATLAS search for particles decaying near the muon spectrometer~\cite{ATLASMS} can place upper bounds on the LSP decay length for sufficiently light squarks. To estimate the bounds on the squark-neutralino model, we assume the same signal efficiency as in the case of a 140 GeV Higgs decaying to two 20 GeV long-lived particles considered in~\cite{ATLASMS}, due to the similar boosts of the long-lived particles in both cases. Scaling the cross-section then gives an upper bound of $\sim 1 \m$ on the neutralino decay length if the squarks are $450 \GeV$. Cross-sections smaller than $\sim 1 \pb$ are not constrained in the models analyzed in~\cite{ATLASMS}, so we do not obtain any decay length bounds for squark masses greater than $\sim 550 \GeV$.

We emphasize that our reinterpretation of the above displaced decay searches is crude and the resulting bounds are not definitive. They do however indicate that, for topologies producing leptons at some level, searching directly for displaced vertices can be a much more powerful probe of these models than conventional missing energy searches. In section~\ref{sec:discovery} we discuss possible approaches for displaced vertex searches that do not rely on lepton triggers, which could provide sensitivity for neutralino decays to $\nu b \bar{b}$.

We have shown that a neutralino LSP in our scenario with a decay length between $\mm - \m$ significantly reduces constraints on superpartner masses (figure~\ref{fig:ChiDecayLength}). This allows squarks as light as 450 GeV, with a heavy gluino $\gtrapprox$ 1.5 TeV (figure~\ref{fig:ChiSquarkGluino}).

\section{Sfermion LSPs}
\label{sec:Constraints}

In this section we discuss the constraints on displaced decays of various possible sfermion LSPs. We estimate the constraints from the displaced decay searches of section~\ref{sec:DisplacedCollider} and also discuss the constraints for each LSP in the promptly-decaying and stable limits, which give the lower and upper bounds respectively on the decay length. We distinguish between ``direct'' bounds that depend only on the existence of a light LSP (possibly light enough to dominate the Higgs width), versus bounds on the production of squarks and gluinos decaying to the LSP. For simplicity, we take the gluino to be 2 TeV throughout this section and consider the bounds on the squark mass, taking all squarks to be degenerate.

\subsection{Sneutrino LSPs}
\label{sec:SnuLSP}

The sneutrinos can be the lightest superpartners for small left-handed slepton soft masses because of the negative $D$-term contributions to their masses. For flavor-universal soft masses the three sneutrinos will be essentially degenerate; we will assume that the three sneutrinos are co-LSPs. 
With bilinear $R$-parity violation the dominant decay mode of the sneutrinos is to $b\bar{b}$. The three sneutrinos may have different lifetimes depending on the relative sizes of the $\epsilon_i$; for simplicity we will assume that these are all equal when discussing sneutrino LSPs. 

The decay lifetime of a sneutrino LSP is 
\be
c \tau = \frac{1}{m} \frac{16\pi}{N_c} \left(\epsilon y_b \sec \beta \right)^{-2} = 11 \text{ cm } \left( \frac{50 \GeV}{m} \right) \left( \frac{10^{-6}}{\epsilon \sec \beta} \right)^2
\ee
where $y_b$ is the Standard Model bottom Yukawa coupling. 

\subsubsection{Direct constraints}
\label{sec:SnuDirect}

\begin{figure}
\includegraphics{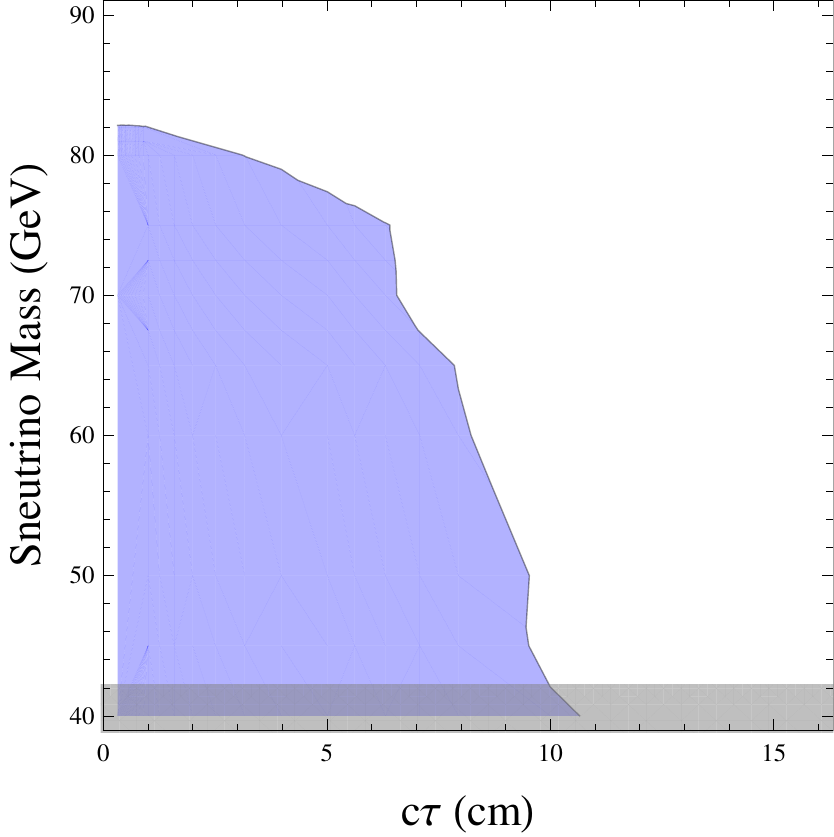}
\caption{Constraints in the mass-lifetime plane on sneutrino LSPs decaying to $b \bar{b}$, assuming that the three sneutrinos are degenerate and have the same lifetime. Short decay lengths are constrained by a LEP four-jet search~\cite{ALEPH4jet} (blue region). Sneutrino masses below 42 GeV are constrained by the total decay width of the $Z$ boson (gray region).
}         
\label{fig:SnuDirect}
\end{figure}

An ALEPH search in four-jet final states constrains the sneutrino mass to be greater than 80 GeV assuming prompt decays~\cite{ALEPH4jet}. For most analyses at ALEPH, charged tracks that do not point back to a cylinder of radius 2 cm and length 20 cm centered at the collision point are ``ignored"~\cite{ALEPHdetector}; assuming that all four jets must point back to this region in order for an event to be accepted by the search, we find that this search excludes the region shown in figure~\ref{fig:SnuDirect} in the sneutrino mass-lifetime plane, namely decay lengths $\gtrsim 10 \cm$, corresponding to $\epsilon_i \sec \beta \lesssim 10^{-6}$. 

A Higgs decay to two sneutrinos will also produce a $4b$ final state. Searches for Higgs decays to two neutral particles that decay promptly to $4b$ have been performed by various LEP experiments~\cite{LEPHiggsReview}. If $m_h \approx 2m_{\tilde{\nu}}$, the sneutrinos from Higgs decays will be slowly-moving and their decays may appear prompt; however because the efficiencies of the $4b$ searches are typically low in this limit this does not place a constraint for proper decay lengths $> 10 \cm$.

\subsubsection{Colored sparticle production}
\label{sec:SnuColored}

The decay modes of the squarks, and thus the bounds on their masses, depend on the spectrum of superpartners with mass between the squarks and the sneutrino LSP. In particular, a sneutrino LSP must be accompanied by a light left-handed charged slepton, with mass determined by the sum rule $m_{\tilde{e_L}}^2 - m_{\tilde{\nu}}^2 = -\cos(2\beta)m_W^2$. Squark cascade decays involving charged sleptons can give rise to events with multiple prompt leptons, which are strongly constrained by LHC searches. Light squark masses are only possible if the branching ratio to charged sleptons is small, for example if the squarks decay dominantly to a neutralino that is lighter than the charged sleptons but heavier than the sneutrino. 

Neutrinos are produced in the decays of neutralinos to sneutrinos, so the ATLAS searches in jets and MET~\cite{ATLASjetsMET1.04,ATLASjetsMET35}, $b$-tagged jets and MET~\cite{ATLASbjet36,ATLASbjet830} and multijets and MET~\cite{ATLASmultijet5,ATLASmultijet1.3} can all have some signal efficiency, just as in the case of a neutralino LSP. The missing energy from the neutrinos is suppressed if the mass splitting between the neutralino and sneutrino is small, as is necessary to prevent decays to charged sleptons. We find that for a spectrum with a 50 GeV sneutrino and an 80 GeV neutralino, the squark mass can be as low as $\sim 500 \GeV$ (for a 2 TeV gluino) if the sneutrino decay is displaced, with the ATLAS multijet plus MET search giving the strongest constraint. If the sneutrino decays promptly, then searches in $b$-tagged final states are sensitive and require the squark masses to be greater than $\sim 550 \GeV$.  

As in the case of a neutralino LSP, an upper bound on the sneutrino LSP lifetime can be determined by considering events in which both sneutrinos escape the detector. In the limit of zero splitting between the neutralino and the sneutrino LSP, the kinematics are identical to the neutralino LSP case. Thus the bounds of figure~\ref{fig:ChiDecayLength} on squark mass as a function of LSP lifetime from ATLAS searches in jets and missing energy also apply to the case of a 50 GeV sneutrino LSP nearly degenerate with a neutralino.
 
\subsection{Stau LSP}
\label{sec:StauLSP}

Renormalization group evolution from a high scale generally causes the right-handed stau to be the lightest sfermion, and possibly the LSP. Bilinear $R$-parity violation predicts a right-handed stau LSP to decay through $LLE$ to a charged lepton and a neutrino, with a $50\%$ branching ratio to taus and $25\%$ each to electrons and muons assuming flavor-universal $\epsilon_i$ (even if the stau has some left-handed component, decays through $LQD$ will be CKM suppressed if the stau is lighter than the top). Since the tau usually produces one charged particle in its decays, the resulting signature is usually either a single large impact parameter track or two incomplete tracks joined at a kink, depending on whether or not the stau track is reconstructed. We will assume that in either case the leptons from stau decay will fail the track quality cuts of LHC searches. The ATLAS search for displaced vertices~\cite{ATLASdisplacedjet} has negligible efficiency for stau decays, as it requires displaced vertices to have at least four outgoing tracks (the branching ratio for tau decays to five or more charged particles is $10^{-3}$). The observability of kinks from stau decays is explored in~\cite{StauKink}; in this section we consider the constraints from existing searches on the stau and squark masses. 

$R$-parity violating decays of the staus have been searched for by ALEPH~\cite{ALEPH4jet}, requiring the stau mass to be $\gtrsim 80 \GeV$ if it decays promptly. Displaced decays through $LLE$ are constrained by the ALEPH searches for kinked or large impact parameter tracks~\cite{ALEPHsleptons, ALEPHgmold, ALEPHgmnew}, giving a stau mass bound near the LEP kinematic limit, $m_{\tilde{\tau}} \gtrsim 100 \GeV$. 

As in the case of sneutrino LSPs, we will consider simplified spectra with degenerate squarks, one light neutralino and a stau LSP. In particular, there are strong constraints on the production of other sleptons (selectrons or smuons) in the cascade decays, as this can give events with multiple prompt electrons or muons. A search in events with missing energy and two or more hadronically-decaying taus~\cite{ATLAS2tau} places the strongest constraints on this scenario. The bounds depend on the stau-neutralino mass splitting; if this is small then the acceptance of taus is reduced. For a stau at 100 GeV and a neutralino at 120 GeV, the squark masses must be $\gtrsim 450 \GeV$ (for a 2 TeV gluino), due to constraints from both jets plus MET and tau plus MET searches (assuming a constant 60\% identification efficiency of hadronic tau decays). If the neutralino mass is increased to 200 GeV however, the tau plus MET search constrains the squark masses to be $\gtrsim 750 \GeV$. 

We note that similar phenomenology can be achieved in $R$-parity conserving models with a stau NLSP decaying to a tau and a gravitino (or other light LSP). Since this decay mode produces fewer muon candidates and more missing energy, the bounds from 0-lepton searches are tighter, requiring squark masses $\gtrsim 850 \GeV$ if the neutralino is 200 GeV and the stau 100 GeV. 

\subsection{Squark LSPs}
\label{sec:SquarkLSP}

Squark LSPs can decay through $LQD$ to a quark and a charged or neutral lepton. If the decay is displaced, then charged leptons will usually not be identified and searches for leptoquarks are not sensitive. The ATLAS search for displaced vertices~\cite{ATLASdisplacedjet} will apply however; using the same methods as in section~\ref{sec:ChiLSP} we estimate the bound on the SUSY production cross-section to be $\sim \pb$ for decay lengths of a few mm, corresponding to a squark mass bound of roughly 500 GeV if the gluino is 2 TeV. As in the case of a neutralino LSP, these bounds would likely rapidly improve as more data is used; if zero events were observed in $1 \fb^{-1}$ then the squark mass bound would increase to $\sim 850 \GeV$.

If the LSP is a mostly left-handed sbottom, however, it can decay dominantly to a bottom and a neutrino. At colliders this will be essentially indistinguishable from decay to a stable massless neutralino, so if the squark decay is prompt then searches in jets and missing energy would apply as usual and the results of~\cite{ATLASjetsMET5} constrain the squark masses to be $\gtrsim 1.4 \TeV$. If we assume, as we have throughout, that jets from displaced decays are accepted by the ATLAS searches in jets and MET, then these constraints apply even if the squark decays are displaced. 

Some ATLAS analyses (such as~\cite{ATLASbjet2,ATLAS1lep5}) explicitly state that the leading jets are required to have associated charged tracks. If such cuts are applied to the jets and MET searches as well, then the bounds on squark masses can be drastically lower. If the squarks are somewhat degenerate, then prompt jets from cascade decays will usually be softer than those from the LSP decay, resulting in the event being rejected. Such a spectrum can be achieved if all of the squark soft masses are equal at some low scale; the effects of running and Yukawa couplings can then make the left-handed sbottom slightly lighter than the other squarks. Similar phenomenology can also be achieved in $R$-parity conserving models with a squark NLSP decaying to a gravitino or axino.

The ATLAS and CMS searches in jets and missing energy could possibly have very low efficiency for these spectra; in the extreme limit they may place no bounds at all. The Run II Tevatron searches in jets and MET~\cite{CDF2jetsMET,D02.1jetsMET} explicitly require the leading jets to have tracks pointing back to the primary vertex and will not apply to this spectrum. Ultimately the most constraining searches may be the Tevatron searches for displaced vertices~\cite{CDFHV,D0displacedjet} or the Run I Tevatron searches~\cite{CDFjetsMET92,D0jetsMET95}, giving lower bounds of less than $200-300 \GeV$ for the squark masses. Of course, the viability of this scenario entirely depends on the hypothesis that LHC searches for SUSY events are not sensitive to events in which the leading jets are displaced.

\section{Novel Higgs Decays}
\label{sec:Higgs}

In the preceding sections we showed that the LSP can be as light as $\sim 40 \GeV$ in most cases if its decays are displaced. This opens up the possibility that the Higgs boson decays into two LSPs. If this decay is kinematically allowed, then it can easily be an important or dominant mode, as it proceeds through gauge couplings rather than Yukawa couplings. 

For sfermions, even LSPs with mass very close to the kinematic limit can dominate the Higgs width, due to the large coupling and lack of phase space suppression in the Higgs decay matrix element. The presence of a $\sim 125 \GeV$ Higgs with Standard-Model-like decays, as suggested by recent LHC searches~\cite{ATLASHiggs,CMSHiggs}, would therefore require all sfermions to be heavier than $\approx 62 \GeV$. 

The Higgs decay rate to neutralinos will depend on the neutralino mixing parameters. A mostly bino LSP with a higgsino mixing angle $\sim .3$ can dominate the Higgs width while giving a small contribution to the $Z$ width and having a small production cross-section at $e^+e^-$ colliders, as observed in~\cite{KaplanUDD}. A neutralino with smaller Higgsino component can give subdominant contributions to the Higgs decay width, possibly allowing for the observation of the Higgs both in standard channels and in events with displaced decays.

If decays to unstable LSPs dominate the Higgs decay width then the LEP bound of 114 GeV on the Higgs mass no longer applies. Because of the highly displaced decays searches in other channels (such as four bottoms \cite{LEPHiggsReview}) will also be insensitive. The only applicable  mass bound then is then the decay-mode independent bound of 82 GeV~\cite{OPALHiggs}. A hidden light Higgs boson would eliminate the little hierarchy problem of the MSSM, in which the apparent heaviness of the Higgs requires large stop masses. Since the collider constraints on squark masses are also greatly weakened in the case of displaced decays of LSPs, a fully natural SUSY spectrum can be achieved in this scenario.

Of course, if the Higgs mass is greater than 114 GeV, it cannot be accommodated into the MSSM without pushing the stop to masses $\gg$ TeV. The uncanceled quadratic contributions from the heavy stop to the electroweak vev introduces a new tuning problem, namely, the little hierarchy problem~\cite{Barbieri:2000gf, Giudice:2006sn, Raby:2007yv, Dermisek:2009si}. A variety of solutions have been proposed to address this problem~\cite{Dermisek:2009si, Graham:2009gy, Martin:2009bg, Choi:2005hd, Kitano:2005wc,  Chacko:2005ra, Ellis:1988er, Espinosa:1998re, Batra:2003nj, Maloney:2004rc, Casas:2003jx, Brignole:2003cm, Harnik:2003rs, Chang:2004db,  Delgado:2005fq,  Birkedal:2004zx, Babu:2004xg, Choi:2006xb}. The lack of experimental observation of light squarks however reintroduces tuning in many of these solutions.  However, bilinear $R$-parity violation can be used in concert with such solutions to alleviate the bound on the squark masses and permit a fully natural solution to the hierarchy problem.

\section{Other Constraints}
\label{sec:BRPVLowEnergy}

Bilinear $R$-parity violation explicitly violates lepton number conservation, and UV models that implement BRPV can also violate baryon number. In this section we discuss the resulting constraints on BRPV from low-energy physics. Neutrino masses are the most sensitive probe of lepton number violation from BRPV, while proton decay constrains the combination of lepton and baryon number violation. We find that these constraints can be satisfied if the $R$-parity violation is weak enough to give displaced vertices.

\subsection{Neutrino masses}
\label{sec:numass}

The neutrino masses induced by BRPV, with all lepton number violation originating from the supersymmetric $LH$ term, are calculated in~\cite{Numass}. The linear combination of neutrinos parallel to $\epsilon_i$ in flavor space will receive the largest contributions to its mass, proportional to $\epsilon^2 \equiv \sum \epsilon_i^2$, from both the sneutrino vev induced by mixing with the Higgses and from a bottom/sbottom loop. Other linear combinations will receive contributions from lepton/slepton loops and will be suppressed by the lepton Yukawas, making them negligible in comparison. We require that the contribution to the largest neutrino mass be less than the square root of the smallest known neutrino mass squared difference, $\sim 10^{-2} \eV$.

The calculations of~\cite{Numass} apply directly to the BRPV scenario we consider. The contribution to the mass of the neutrinos in the $\epsilon_i$ direction is 
$$\Delta m = \Delta m_{tree} + \Delta m_{loop}$$
$$\Delta m_{tree} = \left( \frac{g_1^2}{4M_1} + \frac{g_2^2}{4M_2} \right)\langle\tilde{\nu}\rangle^2$$
$$\Delta m_{loop} = \epsilon^2 \frac{\lambda_b^4}{32\pi^2\tan\beta} \frac{\mu v^2}{\tilde{m}_b^2} \cos 2\beta $$
The sneutrino vev $\langle\tilde{\nu}\rangle^2 \equiv \sum \langle\tilde{\nu_i}\rangle^2$ arises from the supersymmetry-breaking mixing of the sneutrino with the Higgs scalars from the terms $B_{L_i} \mu \tilde{L}_i \tilde{H}_u + \tilde{m}_{H_d,L_i}^2 \tilde{L}_i\tilde{H}_d^* + \hc$:
\be 
\langle\tilde{\nu}_i\rangle = (B_{L_i} \mu \sin\beta + \tilde{m}_{H_d,L_i}^2 \cos\beta) \frac{v}{\tilde{m}_{\nu_i}^2}  
\ee

If we assume that the supersymmetric $\epsilon LH$ operator is the only source of lepton number violation, then we expect the soft parameters $B_{L_i}$ and $\tilde{m}_{H_d,L_i}^2$ to be suppressed by at least a factor of $\epsilon_i y_b^2$ compared to the soft masses $B \mu$ and $m_{H_d}^2$, as generating them requires the $\epsilon_i y_b L_iQD$ term in order to break lepton number and the $y_b H_dQD$ term to break a chiral symmetry of the down quarks. With this suppression the sneutrino, squark and stau LSPs are unconstrained by neutrino masses for the decay lengths of interest to us ($\gtrsim \mm$). The three-body decay of a neutralino LSP can be highly suppressed depending on the squark and slepton masses, possibly requiring relatively large $\epsilon$ to give decay lengths less than $\sim 10 \cm$. We therefore determine the maximal value of $\epsilon$ allowed by neutrino mass constraints and use this to determine the minimum decay lengths of the neutralino for various neutralino and squark/slepton masses.

\begin{figure}
\includegraphics{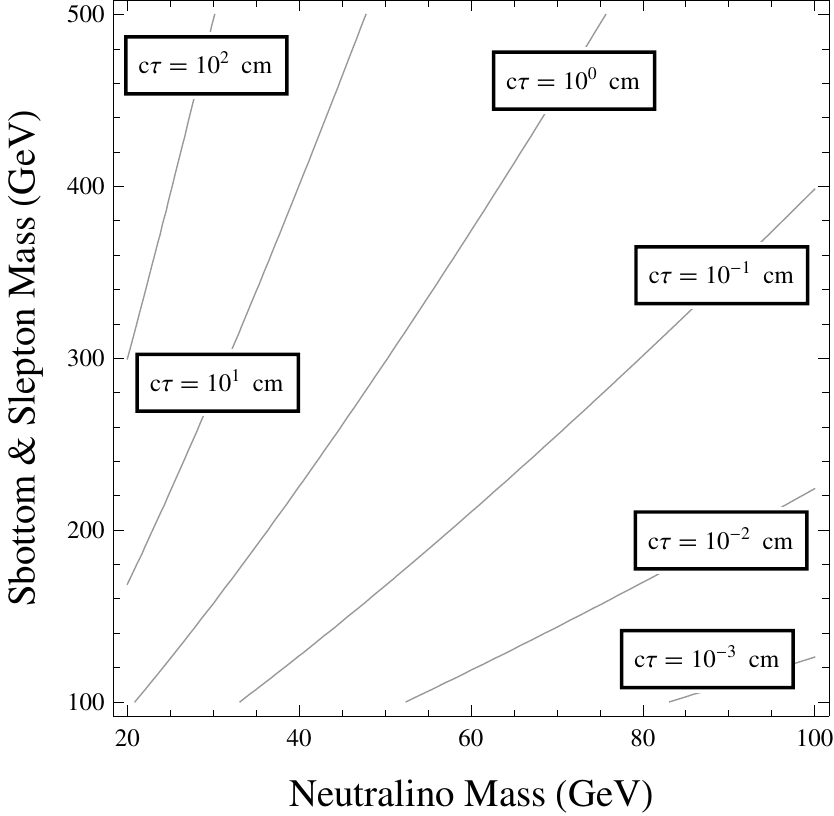}
\caption{Contours of the decay length of a neutralino LSP as a function of the neutralino mass and the masses of the sbottoms and sleptons (assumed degenerate for simplicity), for $\epsilon = 10^{-3}$ and $\tan \beta = 5$. These approximately represent the minimum allowed neutralino decay lengths as this is roughly the maximum $\epsilon$ allowed by neutrino mass constraints. The dominant decay mode in this case is to $\nu b \bar{b}$, through off-shell sbottoms and sneutrinos.
}         
\label{fig:NeutralinoDecayLength}
\end{figure}

The tree-level contribution to neutrino masses from bilinear RPV can be suppressed by large sneutrino masses or small soft lepton-number-violating parameters $B_{L_i}$ and $\tilde{m}_{H_d,L_i}^2$. The latter can be accomplished with low-scale supersymmetry breaking with the soft $L$-violating parameters set to zero at the SUSY breaking scale, as discussed in~\cite{Numass}. At the weak scale these parameters will then be proportional to the squark and gluino soft masses but suppressed by a loop factor in addition to the $\epsilon_i y_b^2$ factor. If we set all of the $A$-terms and soft $L$-violating terms to zero at 10 TeV, take the squarks to have mass $500 \GeV$, the gluino $2 \TeV$, and the sleptons and other gauginos $200 \GeV$, then requiring the neutrino masses from BRPV to be less than $10^{-2} \eV$ implies $\epsilon \lesssim 10^{-3}$ for $\tan \beta = 5$. The resulting decay length of the neutralino is plotted in figure~\ref{fig:NeutralinoDecayLength} as a function of the neutralino and sfermion (sbottom and slepton) masses. We find that neutralino masses $< 100 \GeV$ and sfermion masses of hundreds of GeV can give decay lengths on the order of detector scales (mm $ - \m$), in the range that can hide supersymmetry and give displaced vertex signatures.

\subsection{Proton decay}
\label{sec:pdecay}

The bilinear RPV operator $LH$ only violates lepton number, though UV completions of BRPV models may generate the baryon-number violating operator $UDD$ as well. In this section we consider the constraints on such operators from searches for proton decay.

The strongest proton decay constraints in BRPV scenarios come from limits on the decay mode $p \to \nu K^+$, as this involves second-generation Yukawa couplings where $p \to e^+ \pi$ involves first-generation couplings. The rate for this decay through a right-handed strange squark is approximately 
$$\Gamma_{p \to \nu K^+} \approx \frac{1}{8\pi} \left(\lambda^{''}_{112} \epsilon_i y_s \sec \beta\right)^2 \frac{\Delta m^5}{\tilde{m}_{s_R}^4} $$
where $y_s$ is the Standard Model strange quark Yukawa and $\Delta m$ is the proton-kaon mass difference. We have neglected hadronic matrix elements, so we expect the following bounds to be over-conservative. The current $90 \%$ C.L. bound on the proton lifetime for this decay mode is $6.7 \times 10^{32}$ yr~\cite{SuperKptonuK}, requiring 
$$\lambda^{''}_{112} < 7 \times 10^{-21} \left( \frac{\tilde{m}_{s_R}}{300 \GeV} \right)^2  \left( \frac{10^{-3}}{\epsilon_i} \right) \left( \frac{5}{\sec \beta} \right).$$ 
Weaker bounds will apply for the other $\lambda^{''}_{ijk}$. 

We consider two possible origins for $\lambda^{''}$ in UV completions:
\begin{itemize}
\item With SU(5) matter unification, a $D H^C$ term is produced along with $LH$, which after diagonalization gives $UDD$ terms $\lambda^{''}_{ijk} = \epsilon_k \frac{\mu}{M_{H^C}}y^d_{ij}$ where $M_{H^C}$ is the mass of the colored Higgses. The above bound can then be expressed as 
$$\epsilon < 5 \times 10^{-4} \left( \frac{\tilde{m}_{s_R}}{300 \GeV} \right)^2 \left( \frac{5}{\sec \beta} \right) \left( 10^{-14} \frac{M_{H^C}}{\mu} \right).$$
The colored Higgs mass is typically on the order of the GUT scale, though in some models it can be parametrically heavier or even totally absent from the spectrum.
\item If $R$-symmetry is broken in a hidden sector and communicated to the Standard Model by higher-dimension operators suppressed by a scale $\Lambda$, then $UDD$ operators can be generated with $\lambda^{''}=\frac{\mu_L}{\Lambda}=\frac{\mu}{\Lambda}\epsilon$. The above bound can then be expressed as 
$$\epsilon < 3 \times 10^{-12} \left( \frac{\tilde{m}_{s_R}}{300 \GeV} \right) \left( \frac{5}{\sec \beta} \right)^{1/2} \left(\frac{\Lambda}{\mu}\right)^{1/2}.$$
or
$$\left(\frac{\Lambda}{\mu}\right) > 10^{17} \left( \frac{\tilde{m}_{s_R}}{300 \GeV} \right)^{-2} \left( \frac{\sec \beta}{5} \right) \left(\frac{\epsilon}{10^{-3}}\right)^{2}$$
\end{itemize}   

\section{Discovering Displaced Vertices}
\label{sec:discovery}

We have shown that many existing supersymmetry searches are not sensitive to models with long-lived LSPs, often because they explicitly veto events with displaced decays. However, analyses that specifically search for displaced decays can have even greater discovery potential due to the low Standard Model backgrounds for these types of events (these typically include cosmic rays, scattering in the detector material and artifacts of the detector response). If the datasets for such searches are collected using a specialized trigger for displaced decays, then relatively model-independent bounds on long-lived particles could be obtained; triggering on other aspects of the event (the approach used so far by ATLAS~\cite{ATLASdisplacedjet} and CMS~\cite{CMSdilepton}) can give constraints that are more spectrum-dependent. In this section we discuss both of these strategies in the context of various LSPs and production mechanisms.

The LHCb experiment is potentially an ideal place to discover new displaced vertices over a significant range of decay lengths.  The experiment's primary purpose is to detect rare $b$-quark decays and $B$-meson oscillations by looking for bottom quark production in the far-forward direction, allowing for better reconstruction of multiple vertices and better vertex resolution.  The LHCb vertexing region (or VELO) is nearly a meter long, and the expected resolution is on the order of microns (transverse).  The collaboration can use these capabilities to search for displaced vertices with large track multiplicities with large enough invariant masses.  While the acceptance region is small (on the order of a few percent of hermetic detectors), it is made up by the minimal trigger requirements -- events with a 2.1 GeV $p_T$ muon or 3.6 GeV $p_T$ pass ``level-zero'' triggers, and displaced vertices can even be triggered on at the hardware level. In an unreleased analysis of 2010 data LHCb has already searched for a Higgs decaying into two neutralinos which decay with displaced vertices into multiple jets~\cite{ABay}.

The ATLAS collaboration has developed triggers for decays in the calorimeters and muon systems~\cite{ATLASLLTrigger}, at a distance from the beamline of order meters. These include triggers on 
\begin{itemize}
\item trackless jets with a trackless muon in the jet cone, sensitive to decays beyond the inner regions of the tracking detector,
\item trackless jets with much more energy deposited in the hadronic calorimeter than in the electromagnetic calorimeter, sensitive to decays within the calorimeters, and
\item multiple clustered tracks in the muon spectrometer isolated from inner detector tracks in angle, for decays in the muon system or the outer parts of the hadron calorimeter. This is used for the Higgs decay search described in section~\ref{sec:DisplacedJets}.
\end{itemize}
For the appropriate decay distances these triggers will all have some efficiency for sneutrino and neutralino LSPs, as the decays to bottom quarks produce jets and sometimes muons ($\sim 20\%$ branching ratio per $b$). Decays in the outer silicon layers ($\sim 30-50 \cm$) will appear trackless~\cite{ATLASLLTrigger}, so these triggers will be efficient for LSP boosted decay lengths ($\gamma c \tau$) on the order of $\sim 30\cm$ or more. None of these triggers are applicable for stau or squark LSPs (the latter will generally be associated with charged particles from showering even if the $R$-hadron is neutral).

In the absence of appropriate displaced decay triggers, searches must rely on conventional triggers such as those for leptons, missing energy and high-$p_T$ jets. All of the bilinear-RPV-induced decays we consider include neutrinos in the final state, so missing energy triggers can be efficient. An analysis similar to the ATLAS search for displaced vertices~\cite{ATLASdisplacedjet} using missing energy, hard jet and/or multijet triggers could potentially provide excellent sensitivity for the case of neutralino, sneutrino or squark LSPs. Even in more general displaced decay models without final-state neutrinos, missing energy triggers can be efficient for longer decay lengths due to the spurious MET introduced by mismeasurement of jet momenta.

In the case of the stau LSP decaying through $LLE$, the signature is generally a kinked track rather than a displaced vertex (and very rarely a displaced vertex with more than three tracks). This scenario therefore requires different search strategies~\cite{StauKink}. The stau case can also give more objects to trigger on: most events will have prompt taus, and many will have displaced muons. 

\section{Conclusions}

We have argued that current collider searches leave a general ``hole'' in sensitivity to particles with decay lengths on the scale of the detector size, $\sim \mm-\m$. The products of these decays are often not properly reconstructed in standard searches that assume prompt decays. A limited number of analyses (including~\cite{ATLASdisplacedjet,ATLASMS,ATLASdisaptrack,CMSdilepton,CMSdispphoton} at ATLAS and CMS) specifically search for displaced decays of new particles, though these generally require specific objects to trigger on and as of yet do not cover the full range of possibilities for displaced decay topologies.     

We illustrated this hole with a simple, well-motivated supersymmetric model with bilinear $R$-parity violation.  By adding a new sector that breaks $R$-parity one dominantly generates the bilinear terms in the superpotential $W \supset \mu_L L H_u$. The coefficients can be naturally small and decays are further suppressed by Yukawa couplings, leading to long displaced vertices in every SUSY event. The decay modes are largely determined by the Standard Model Yukawa couplings and typically involve bottom quarks, taus and neutrinos. This model naturally avoids proton decay problems by generating only lepton number violating operators without an arbitrary choice of which $R$-parity violating coefficients to turn on.

We have shown that this form of $R$-parity violation with displaced decays can allow all squarks to be as light as $450 \GeV$ for a light neutralino LSP (figures~\ref{fig:ChiSquarkGluino} and~\ref{fig:ChiDecayLength}), far lighter than is possible in the simplest scenario of stable light neutralinos. 
This is achieved because of the suppression of missing energy due to LSP decay (despite the presence of neutrinos in the decay) and the displacement of the LSP decay preventing reconstruction of $b$-jets and leptons. The strongest constraints come from ATLAS searches in jets and MET in $4.7 \fb^{-1}$~\cite{ATLASjetsMET5,ATLASmultijet5}, with selections requiring high jet multiplicity forcing the gluino mass to be large, $\gtrsim 1.5 \TeV$ if the squarks are light. Such a gluino-squark hierarchy can be achieved with very low scale ($\sim 5 \TeV$) SUSY breaking or in models with Dirac gauginos or negative high-scale squark masses. For other LSPs light squark masses may also be achieved for certain SUSY spectra.   

The possibility of light squark masses in this scenario directly eases the fine-tuning of the weak scale. Furthermore, for an LSP lighter than half the Higgs mass, the Higgs can decay dominantly into two LSPs.  This would evade existing Higgs searches and allow the Higgs to be as light as $\sim 100$ GeV, which would remove the little hierarchy problem of the MSSM.  Of course, for heavier LSPs the Higgs decays as in the $R$-parity conserving MSSM and the standard searches apply. In the case of neutralino LSPs the Higgs branching ratio to LSPs can be subdominant, making displaced vertices an additional channel in which to observe a Higgs with otherwise SM-like decays.

This scenario can allow for a natural, light spectrum of superpartners. Furthermore it would imply a large signal cross-section (up to $\approx 2 \pb$) for searches targeting displaced decays on detector length scales. The LHCb detector may be ideally suited to directly triggering on highly displaced vertices. Searches at ATLAS and CMS may have to rely on identifying displaced vertices in data samples collected using other trigger objects, such as missing energy, or on looking for decays in the calorimeters or muon systems.  

Some $R$-parity conserving models can give signals similar to those considered here if they contain LSPs that are very weakly coupled to the rest of the spectrum. Displaced decays of stau or squark NLSPs to light gravitinos give signatures similar to those in the bilinear RPV scenario. Another possibility is a right-handed sneutrino LSP; models in which the masses and couplings of the right-handed neutrinos are generated by supersymmetry breaking~\cite{ArkaniHamed:2000bq,ArkaniHamed:2000kj,Borzumati:2000mc,Borzumati:2000ya,MarchRussell:2004uf} can explain the smallness of neutrino masses and give the right-handed neutrinos small couplings that lead to displaced vertices.   

Naturalness is coming into increasing tension with collider constraints.  The hierarchy problem strongly suggests a new colored particle around the top mass but current limits require such a particle to be so heavy that some tuning is still required.  There are only a few possible theoretically well-motivated remedies that still allow light, colored particles including squeezing parts of the spectrum (e.g.~Stealth Supersymmetry \cite{Fan:2011yu}) or allowing the new particles to decay on detector length scales as we have discussed here.

\section*{Acknowledgments}

We would like to thank Kaustubh Agashe, Asimina Arvanitaki, Aurelio Bay, Kyle Cranmer, Savas Dimopoulos, Beate Heinemann, Kiel Howe, Patrick Janot, Christos Leonidopoulos, John March-Russell, Jeremy Mardon, Josh Ruderman, Dan Tovey, Giovanni Villadoro, Gordon Watts and Itay Yavin for useful discussions. P.S. is supported by a Stanford Graduate Fellowship. This work was partially supported by ERC grant BSMOXFORD no. 228169.

\end{document}